# Polarization-diversity based Rotation Sensing Methodology using COTS UHF RFID tags

Florian Muralter*, Fabian Muralter, Hugo Landaluce, Asier Perallos

*Abstract*—Phase-based sensing using ultra high frequency (UHF) radio frequency identification (RFID) has, in recent years, yielded numerous additions to the Internet of Things (IoT). This work presents a polarization diversity based rotation sensing methodology using common-off-the-shelf (COTS) UHF RFID tags identified with a software defined radio (SDR) UHF RFID reader. The proposed methodology uses the tag-to-reader message after fully coherent demodulation to calculate a difference signal of the backscatter load modulation states. This sequence is then used to compute the rotation speed by evaluating its phase change over time. Experimental results are used to validate the theoretical model and to evaluate the performance and limitiations of the proposed system.

## I. INTRODUCTION

THE Internet of Things (IoT) is a paradigm envisioning a plethora of interconnected real-time sensing and actuating devices communicating via the Internet or other telecommunications technologies [1]. Radio frequency identification (RFID) is an automated identification technology and is considered one of the fundamental technologies enabling the IoT [2]. One such RFID system typically consists of at least one reader (interrogator) and a minimum of one tag (transponder) [3]. Connecting the reader to the internet allows the acquired information to be streamed and distributed globally. In recent years, the capabilities of RFID have been extended from being identification only to, e.g., performing sensing tasks or ranging. The proposed RFID sensing applications range from temperature/humidity sensing to structural monitoring, and can be grouped into (a) analog RFID sensing, where the sensed parameter alters the backscattering of the antenna, and (b) digital RFID sensing, where the RFID tag is equipped with an additional sensor and the obtained data is backscattered using the EPC-C1G2 communication protocol's [4] specific commands [5]. If commercially available RFID solutions are used for the development of such platforms, the term common-off-the-shelf (COTS) is used.

A rotation is "a special kind of motion, for which at least one point in space remains at rest" [6]. Measuring the rotation speed or the frequency of the rotation represents a very recent addition to COTS RFID sensing [7], [8]. Nevertheless, the existing solutions do not offer the possibility to measure

Manuscript received __, __.

This work has been part-funded by the Basque Government under the project IoTrain (RTI2018-095499-B-C33).

Fl. Muralter, H. Landaluce and A. Perallos are with DeustoTech (University of Deusto). Fa. Muralter is currently without affiliation.

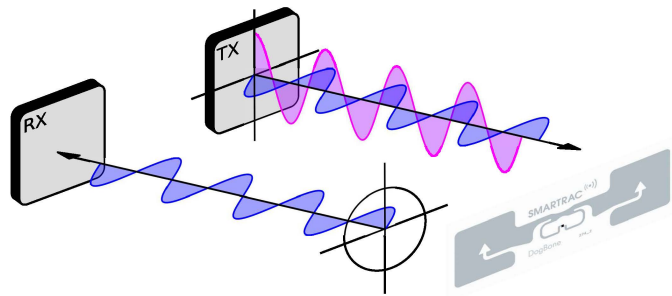

Figure 1: Visualization of the polarization diversity (circular - linear) in an RFID system.

the rotation speed in real-time, nor do they allow for the acquisition of rotation data from more than one tag at the same time.

In this work a methodology for real-time monitoring of the rotation frequency of a COTS Ultra High Frequency (UHF) RFID tag based on the tag-reader polarization diversity (see Fig. 1) is presented. The proposed technology performs one rotation frequency measurement per backscattered Electronic Product Code (EPC) message and, thus, the rotation data can be assigned to the tag's EPC which allows for the acquisition of rotation data from multiple tags within the read range of the interrogator. Considering the Nyquist sampling theorem and the maximum backscatter-link frequency (BLF) of 640 kHz according to the EPC-C1G2, the maximum theoretical rotation frequency that can be sensed by the proposed platform is 320 kHz. This work shows real-time measurements up to 40 Hz with a maximum relative error of less than 10 %, simultaneously identifying three transponders inside the read range.

## II. RELATED WORK

This section presents, on the one hand, the main traditional approaches for measuring the rotation speed. And on the other hand, it presents and analyses the most recent strategies based on RFID COTS for this purpose.

### A. Traditional Approaches

The rotation speed is usually obtained by sensing the angular displacement over time. Typical devices used for this purpose are pulse sensors, specialized sensors and cameras.

- Pulse sensors: send speed data in the form of electric pulses. The number of pulses per revolution and the symmetry of these pulses determine the accuracy of the measurement. Some of the most typical sensors generating electric pulses per rotation are proximity sensors, sensing the teeth on a gear; and the high-resolution version, which are photoelectric or laser sensors, which sense the reflective target. And finally encoders, which combine several transmitter/receiver signals that generate a recognizable pulse sequence used to control position and speed with very high precision. These techniques, however, require a line of sight between the transmitter and the receiver to work properly [9]–[11].

- Specialized sensors: measure variables like centrifugal forces to calculate the speed of rotation (e.g., accelerometers [12]), or the rotational movement (e.g., gyroscopes). However, these sensors require wired connections to the shaft to allow the movement, or wireless sensors which would require the use of batteries, resulting in complex and intrusive solutions. In addition to this, gyroscopes can get saturated with a few revolutions per second.
- Cameras: record high-frame-rate image sequences to measure the frequency of periodic motions [13], [14]. However, these solutions are rarely adopted because of high cost and because cameras require very specific lighting in harsh or changing environments.

The solution proposed in this work, as opposed to the traditional measuring techniques, provides a non-intrusive, contactless methodology without the need of a line-of-sight, to measure the speed of a rotary motion using passive devices.

### B. RFID-based Approaches

Fully coherent demodulation represents the most common demodulation technique used in commercial UHF RFID readers [15]. This technology allows for the acquisition of both signal power and phase [16]. However, due to the decoding technique used for the received backscatter response, the resulting phase presents a 180 degree ambiguity, thus, losing valuable information. Miesen et al. in [17] presented an alternative method allowing for a $2\pi$, full 360 degree ambiguity, but the proposed method is not yet implemented in commercially available UHF RFID readers.

The most common application using the signal phase in UHF RFID is indoor localization [18]–[20], but recent years have seen promising results using phase evaluation techniques for sensing environmental parameters such as the relative humidity (RH) [21], or the temperature [22].

A first approach towards using an RFID system's polarization diversity for sensing a tag's displacement and tilt angle has been proposed in [23]. The method used therein uses a Voyantic Field Engineer Kit [24], to simulate a fully coherent UHF RFID system and to investigate the influence of the displacement and the tilt angle on the signal phase and power. Further investigations on the effect of tag polarization on the resulting signal phase were presented in [25]. However, no model was derived from the empirical results.

In [7], a COTS UHF RFID rotation sensing platform is presented. The proposed method utilizes the phase change introduced due to a change in the tag-reader distance, to obtain the rotation speed. The rotation signal is then reconstructed from one sample per identification using compressive sensing. The described method is limited to the case where an appropriate distance and distance change is present during rotation. Furthermore, $N = 3000$ reads are suggested to provide an accurate estimation of the rotation frequency, which represents a read time of $> 3s$ considering the fastest available UHF RFID readers. [8] represents an improvement of the method previously described, where 2 tags are mounted on a single rotating device. This solution was proposed to decrease the influence of multipath effects and the shaking of the rotating device. Additionally, using the relative phase of the two mounted tags, the influence of the tag-reader distance on the calculated angle is omitted. Nevertheless, this methodology also requires an appropriate distance change during rotation and a large number of reads to reconstruct the rotation signal. A different approach, where both the reader and the tag antenna are linearly polarized, was proposed in [26]. This setup results in a change of the Received Signal Strength Indicator (RSSI) during rotation of the tag. The rotation frequency is then deduced from the peaks in the backscattered power. A similar setup was used in [27] to develop a two-antenna whiteboard (reader) to draw shapes using a pen (tag), considering the orientation of the tag with respect to the reader antennas. Considering a typical read rate of 40 reads per second, considering the Shannon theorem, a maximum rotation frequency of 20 Hz could be detected.

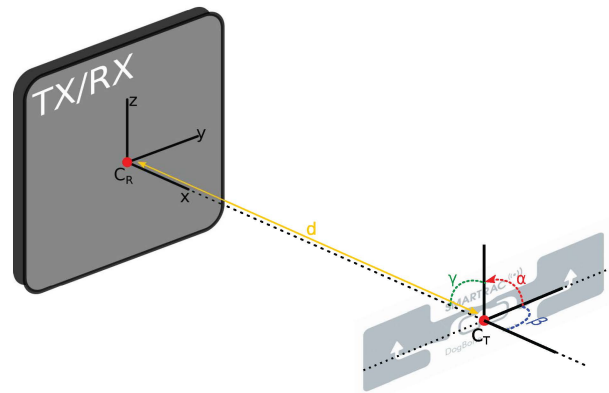

Figure 2: Model of the UHF RFID tag rotating in the field of the reader's TX/RX antenna.

## III. BACKGROUND

The main theoretical concepts necessary for the understanding of the rotation sensing procedure are explained within this section. Some signal theory concepts are analyzed, followed by an explanation of the effects of polarization diversity on the reader-tag radio frequency (RF) communications. Finally, the impact of polarization diversity on the phase of the tag-to-reader communication is explained.

### A. Signal Theory

In UHF RFID the emitted RF signal is typically considered scalar and can be written as

$$s(t, z) = A \cdot sin(2\pi f t - kz + \phi), \qquad (1)$$

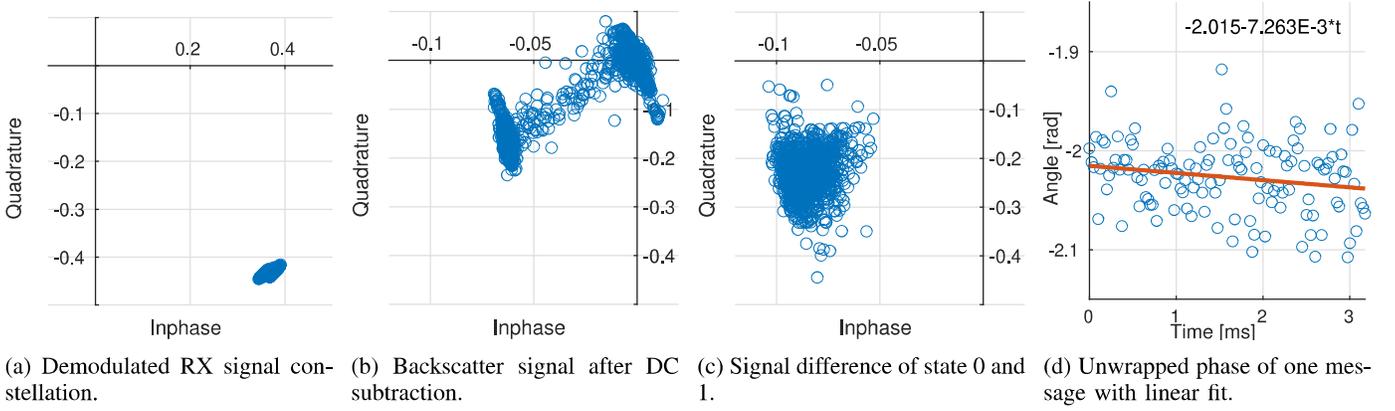

(a) Demodulated RX signal constellation.
(b) Backscatter signal after DC subtraction.
(c) Signal difference of state 0 and 1.
(d) Unwrapped phase of one message with linear fit.

Figure 3: Visualization of the signal processing steps considering the proposed methodology.

where $A$ is the signal amplitude, $f$ denotes the carrier frequency and $\phi$ represents the initial phase value. The term $-kz$ with $k = 2\pi/\lambda$ describes the signal change as a function of the distance considering the propagation direction of the plane wave. This representation is a simplification for the case where the transmitting antenna is linearly polarized [28].

The E-field of a plane electromagnetic (EM) wave propagating in positive z-direction considering polarization can be written as:

$$\begin{pmatrix} E_x e^{j\phi_x} \\ E_y e^{j\phi_y} \end{pmatrix} e^{j(2\pi ft - kz)}, \tag{2}$$

where $E_x$ and $E_y$ represent the amplitudes in x-direction and y-direction, respectively and $\phi_x$ and $\phi_y$ denote the corresponding phase values at $t = z = 0$. Considering this representation, the vector in brackets is called the Jones vector of the polarized EM wave, where for circular polarization the following amplitude and phase relations hold:

$$E_x = E_y, \tag{3}$$

$$\phi_x = \phi_y \pm \frac{\pi}{2}. \tag{4}$$

The sign of the $\pi/2$ phase shift between $\phi_x$ and $\phi_y$ depends on whether the the the circular polarization of the wave is right-handed or left-handed. The Jones Calculus is typically used to calculate the effect of linear optical elements on an electromagnetic wave. [29]

### B. Polarization Diversity

The following section will be used to offer a step-wise explanation of the effects of polarization diversity on the backscattered signal. This calcuation considers a reader antenna with right-hand circular polarization (RHCP) and a linearly polarized (LP) tag antenna being positioned as depicted in the model in Figure 2 with a separation of at least $d = 2\lambda$ to allow for far-field assumptions.

(1) An RHCP EM wave is emitted by the UHF RFID reader antenna with the Jones vector being

$$J_{\mathrm{RHCP}} = \frac{1}{\sqrt{2}} \begin{bmatrix} 1 \\ -j \end{bmatrix}. \tag{5}$$

(2) The linear dipole antenna of a typical UHF RFID tag acts as a linear polarizer on the impinging EM wave. Depending on the tag's tilt angle $\alpha$, the corresponding Jones matrix $M_{\mathrm{LProt}}$ can be calculated as

$$M_{\mathrm{LProt}}(\alpha) = M_{\mathrm{rot}}(\alpha) \cdot M_{\mathrm{LP}} \cdot M_{\mathrm{rot}}(-\alpha), \tag{6}$$

with

$$M_{\mathrm{LP}} = \begin{bmatrix} 1 & 0 \\ 0 & 0 \end{bmatrix}, \tag{7}$$

and

$$M_{\mathrm{rot}}(\alpha) = \begin{bmatrix} cos(\alpha) & sin(\alpha) \\ -sin(\alpha) & cos(\alpha) \end{bmatrix}. \tag{8}$$

$$M_{\mathrm{LProt}} = \begin{bmatrix} cos^2(\alpha) & cos(\alpha)sin(\alpha) \\ sin(\alpha)cos(\alpha) & sin^2(\alpha) \end{bmatrix}. \tag{9}$$

Thus, the resulting Jones vector $J_{\mathrm{LProt}}$ after the linear polarizer can be written as

$$J_{\mathrm{LProt}} = M_{\mathrm{LProt}} \cdot J_{\mathrm{RHCP}}. \tag{10}$$

(3) The RFID reader's RHCP transmitting antenna is considered an LHCP polarizer for the reception path with the corresponding Jones matrix $M_{\mathrm{LHCP}}$ being

$$M_{\mathrm{LHCP}} = \frac{1}{2} \begin{bmatrix} 1 & -j \\ j & 1 \end{bmatrix}. \tag{11}$$

The resulting Jones vector of the received EM wave $J_{\mathrm{RX}}$ can then be calculated as a superposition of the effects of both the linear polarizer (tag antenna) and the circular polarizer (RX antenna)

$$J_{\mathrm{RX}} = M_{\mathrm{LHCP}} \cdot M_{\mathrm{LProt}} \cdot J_{\mathrm{RHCP}}, \tag{12}$$

and simplified to

$$J_{\mathrm{RX}} = \begin{bmatrix} 1 \\ j \end{bmatrix} \cdot e^{(-2j\alpha)}, \tag{13}$$

where the introduced phase term $e^{(-2j\alpha)}$ will be discussed in detail as part of the following section III-C.

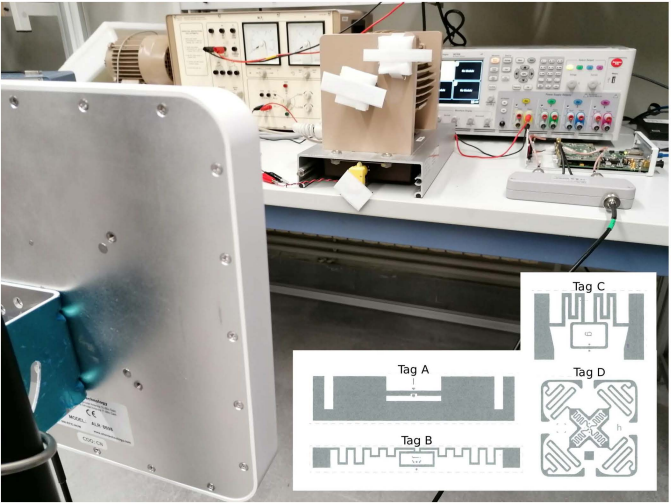

Figure 4: Measurement Setup

### C. Phase Evaluation

The received baseband signal using coherent demodulation $E_{RX}$ can be written as a sum of the components $A$ - Transmit/Receive Leakage, $B$ - Environmental Multipath, $C$ - Antenna Structural Scattering, $D_i$ - Load-dependent Antenna Scattering, and $N$ - Noise [15], [16] resulting in

$$E_{RX} = A + B + C + D_i + N, \qquad (14)$$

where the corresponding phase of $E_{RX}$ is:

$$\phi = \phi_{os} + \phi_{prop} + \phi_{bs} \qquad (15)$$

with

$$\phi_{prop} = -2kd = -2\frac{2\pi f}{c}d. \qquad (16)$$

$\phi_{os}$ represents the signal offset due to cables, reader/antenna components, etc., $\phi_{prop}$ the phase due to the propagation delay, $\phi_{bs}$ the phase shift caused by the load modulation, and $d$ the distance from the RX/TX antenna.

The influence of the polarization on the signal phase can be considered by adding a separate term $\phi_{pol}$, which yields

$$\phi = \phi_{os} + \phi_{prop} + \phi_{bs} + \phi_{pol}, \qquad (17)$$

with

$$\phi_{pol} = -2\alpha \qquad (18)$$

where $\alpha$ denotes the angle of the tag rotation with respect to the x-axis in mathematically positive sense.

## IV. METHODOLOGY

The following section explains the signal processing steps performed to calculate both the phase and the rotation speed of a given tag response.

### A. Phase Calculation

Figure 3a shows the received signal after coherent demodulation.

(1) The DC content of the signal is estimated during the time frame previous to the backscattering of the tag ($T_1$ according to the EPC-C1G2 [4]). This estimate corresponding to the absorbing state of the load modulation is then used to center the received constellation in the complex plane resulting in the data-points visualized in Figure 3b.

(2) The decoding of the received message, as described in [30], is based on the detection of positive and negative edges at the end/start of each FM0 symbol. This information is used to assign the correct load modulation state to each of the data bits.

(3) The backscatter difference signal $b[n]$ is then calculated as the difference of the reflected and the absorbing state at each symbol transition as shown in Figure 3c.

$$b[n] = \begin{cases} s[n-] - s[n+] & \text{if positive edge} \\ s[n+] - s[n-] & \text{if negative edge} \end{cases}, \qquad (19)$$

where the $n-$ refers to the data sample before the signal transition and $n+$ to the data sample after the signal transition. The signal $b[n]$ avoids the influence of all static content contributing to Equation 14. Furthermore, this method allows for the acquisition of equidistant data with the same sampling frequency as the datarate of the backscatter response.

(4) The backscatter signal power and angle can then be calculated as

$$RSSI = \frac{1}{2}\frac{|D_1 - D_0|^2}{Z_0} = \frac{|b[n]|^2}{2Z_0}, \qquad (20)$$

$$\phi = ang(D_1 - D_0) = atan\left(\frac{\mathfrak{Re}\{b[n]\}}{\mathfrak{Im}\{b[n]\}}\right), \qquad (21)$$

where $D_1$ and $D_0$ stand for the complex values representing the reflecting and the absorbing state, respectively. $Z_0$ represents the reference impedance of the system. The resulting signal phase $\phi$ has a $2\pi$ ambiguity as opposed to commercial readers which have a $\pi$ ambiguity [31].

### B. Rotation Frequency Calculation

Considering that the tag-reader distance is constant $d = const.$, the phase of the backscattered signal changes depending on the tag orientation according to

$$\phi(t) = \pm 2 \cdot \alpha(t) + \phi_0, \qquad (22)$$

where $\alpha$ is the tilt angle of the tag as shown in Figure 2 and $\phi_0$ corresponds to the signal phase at $\alpha = 0$.

For a perfectly circularly polarized reader antenna and a perfectly linear dipole tag antenna with no vertical component, a full rotation of the tag would, thus, result in a circle with the center being aligned with the center of the complex plane. The resulting phase change per angle difference is constant. The angular speed $\omega_{rot}$ can then be calculated as

$$\omega = \frac{\Delta\phi}{\Delta t} \qquad (23)$$

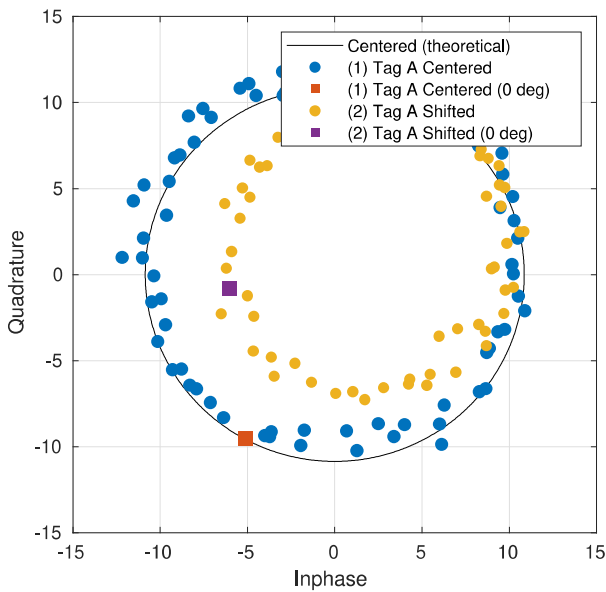

Figure 5: Constellation of backscatter signal vector for a full tag rotation.

where $\Delta\phi$ corresponds to the gradient of the unwrapped phase data and $\Delta t$ to the sample period of the backscatter signal. Considering that the obtained phase values show a Gaussian normal distribution [20], the mean of $\Delta\phi$ is calculated fitting a linear regression to the obtained phase data (see Figure 3d).

## V. Experimental Results

The following section describes the measurements carried out to evaluate the performance of the proposed platform. The measurements were performed in a laboratory setup with an RFID system consisting of a SDR UHF RFID reader based on a USRP N210 using an adapted version of the GNURadio reader software presented in [30] and COTS UHF RFID tags. The USRP N210 is connected to a VSWR bridge to allow for the usage of a single TX/RX antenna. The experimental setup including the 4 different investigated COTS UHF RFID tags is shown in Figure 4, where the corresponding RFID chips are:

- Tag A: Impinj Monza R6
- Tag B: Impinj Monza R6-P
- Tag C: NXP Ucode 8
- Tag D: Impinj Monza 4E

The EPC-C1G2 communication protocol [4] was used with the following parameters:

- Tag-to-reader datarate: 40kHz
- Tag-to-reader encoding: FM0
- PC+EPC message length: 128 bits

### A. Phase Evaluation

Firstly, we performed measurements to investigate the influence of the tag position and orientation on the measured phase. Tag A was positioned at a distance of $0.5\,\mathrm{m}$ ($\sim 2\,\lambda$) in parallel to the reader antenna, with the centers of the tag antenna and the reader antenna being aligned. A full rotation of the transponder was measured in steps of 5 degrees. Figure 5 shows the distribution of the samples on the complex plane. It can be seen, that the resulting constellation represents a circle, with its center at the center of the coordinate system, thus, showing the behaviour predicted in the calculations in section IV-B. The second curve represents a rotation of the tag around its center, where the transponder's center was moved by $-0.2\,m$ on the y-axis according to the model shown in Figure 2. The resulting constellation also represents a circle, but due to the increase in the distance, the radius is slightly smaller. Furthermore, the shape's center now yields an in-phase offset related to the translation on the y-axis.

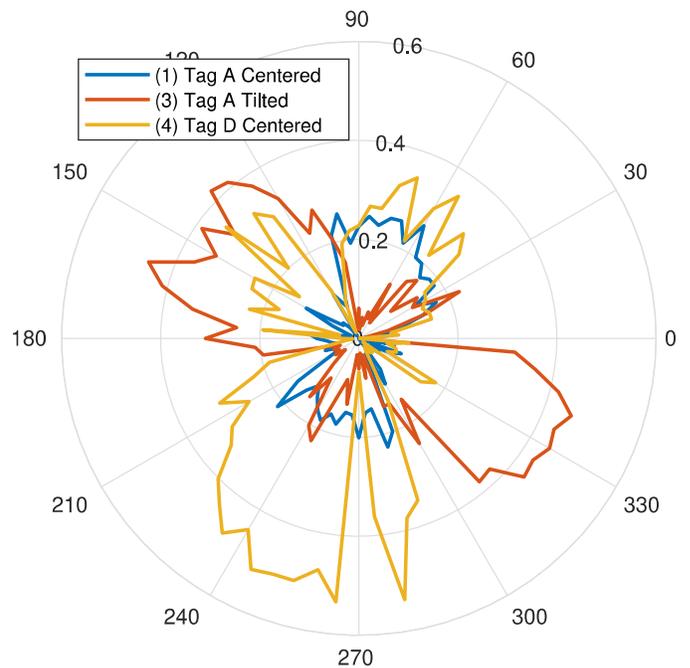

Figure 6: Polar plot of the phase error for a full tag rotation considering different scenarios.

The same measurement has been performed considering four different scenarios:

(1) Tag A center aligned with reader antenna center
(2) Tag A center shifted by $-0.2\,m$ on the y-axis
(3) Tag A tilted ($\beta = 30°$)
(4) Tag D center aligned with reader antenna center.

All measurements were performed with the transponder positioned at a distance of $x = 0.5\,m$ from the reader's TX/RX antenna.

Figure 6 shows the absolute phase error considering $\phi_{pol}$ as the main influence on the backscatter signal phase during rotation, when the tag-to-reader distance $d$ is constant. This parameter is obtained by calculating the absolute value of the difference between the measured signal phase and the model presented in 18. It can be seen that scenario (1), where the centers are aligned, shows the smallest deviation from the theoretical assumptions. The slightly tilted, figure-of-eight shaped error occurs due to the polarization of the reader antenna not being perfectly circular. This behavior can be taken into account in the model by considering an elliptically polarized (EP) reader

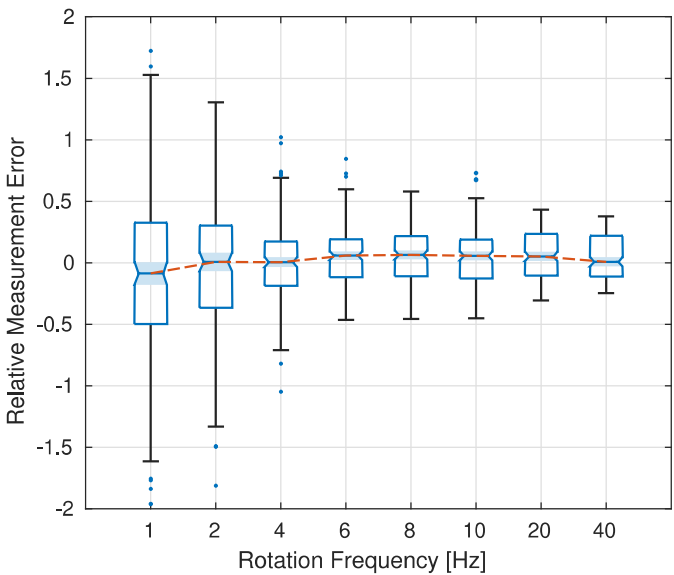

Figure 7: Relative measurement error as a function of the rotation speed.

antenna, substituting the Jones vector $J_{\text{RHCP}}$ by

$$J_{\text{RHEP}} = \frac{1}{\sqrt{A^2 + B^2 + C^2}} \left[ \begin{array}{c} A \\ B - Cj \end{array} \right], \quad (24)$$

and changing the corresponding Jones matrix $M_{\text{LHCP}}$ to $M_{\text{LHEP}}$ accordingly.

For the tilted measurement (3), an increase of the error is visible. However, the behavior of the error as a function of the rotation angle is similar, as the centers of the tag and reader antennas are also aligned in this configuration. The size and rotation of the figure-of-eight shape results from the change in the angle of incidence at the linear polarizer (tag antenna). To consider this effect in the presented model, the model calculus has to be extended to three dimensions. In scenario (3) the rotation axis is perpendicular to the tag plane, but does not match the z-axis, as it is tilted away by the angle $\beta$. This, can only be taken into account using a three dimensional rotation matrix for the tilt, additionally to the rotation matrix $M_{\text{LProt}}$. Measuring the rotation angle $\alpha$ with Tag D yields the results deviating the most from the theoretical description presented in section III, as the double dipole antenna of this transponder design does not represent a purely linear polarizer, and thus, the Jones calculus presented in Section III-B has to be adapted accordingly. In general, taking into account the tag polarization matrix $M_{\text{TP}}$, the rotating tag's Jones matrix can be calculated by

$$M_{\text{TProt}}(\alpha) = M_{\text{rot}}(\alpha) \cdot M_{\text{TP}} \cdot M_{\text{rot}}(-\alpha). \quad (25)$$

### B. Rotation Frequency Evaluation

The following measurements are used to show the functionality and flexibility of the proposed system in addition to the evaluation of its performance and limitations.

*1) Number of Identifications:* At a datarate of $40\,\text{kHz}$, and considering a tag response of 128 bits, the time for estimating the rotation frequency is approximately $3\,\text{ms}$. At a rotation frequency of $1\,\text{Hz}$ this would mean, that during one tag response, the covered angle would be about $1\,°$ or $0.02\,\text{rad}$, which would need a measurement system with very high accuracy to deduce the correct rotation speed from a single measurement. Using the mean of multiple measurements, an increased accuracy can be achieved. The required sample size $N$ can then be calculated as

$$N >= \frac{z^2 \cdot \sigma \cdot (1 - \sigma)}{e^2}, \quad (26)$$

where $z$ represents the Z-score corresponding to a chosen confidence level. $\sigma$ is the standard deviation of the dataset and $e^2$ denotes the error margin. Considering a confidence level of $90\,\%$ and an error margin of $10\,\%$, the mean required sample size calculated using a 200 measurement dataset per evaluated rotation frequency is $N >= 58$. Increasing the desired confidence level to $95\,\%$ and decreasing the error margin to $5\,\%$ results in $N > 328$.

It takes around $5\,\text{ms}$ to identify 1 tag in a common RFID system using EPC-C1G2 protocol with a typical Tari of $6.25\,\text{µs}$ and $40\,\text{kHz}$ tag data rate. The inventory round time increases to $17\,\text{ms}$ if there are 3 tags in the antenna range. Thus, $290\,\text{ms}$ would be needed to obtain an error margin of $10\,\%$, or $1.6\,\text{s}$ to obtain an error margin of $5\,\%$.

*2) Rotation Speed:* To allow for an accurate set-point when measuring the rotation speed of the transponder, a DC motor was used, where the speed of the rotation could be set with a possible range of $[-3000, 3000]$ revolutions per minute (RPM). The transponders were mounted on custom-fabricated foam fixtures to separate the tag from the metal housing of the motor (see Figure 4). Figure 7 shows the relative measurement error as a function of the rotation speed. For each set-point, 200 messages were acquired, and, the corresponding rotation speed was computed. With an increase of the rotation speed, a decrease of the relative measurement error is visible. This behaviour can be explained, considering the fact that a higher rotation speed corresponds to a larger angle covered during a single identification. Thus, the most accurate measurement would be achieved if rotation period matches the time of the backscatter response. It can be seen that for all sets of measurements, the mean relative measurement error is below 10%.

*3) Tag Diversity:* As explained in Section III, the shape of the tag antenna also influences the shape of the constellation resulting from the rotation. The measurement is expected to be more accurate, the closer the tag antenna is to a pure dipole antenna. However, form factor, matching network design and read range are further parameters considered during the design process of a UHF RFID tag [32]. Figure 8 shows the mean relative measurement error as a function of the rotation speed for 4 different tag assemblies. It can be seen that all transponders except for Tag E show a decrease of the measurement error with an increase of the rotation speed. Tag

 

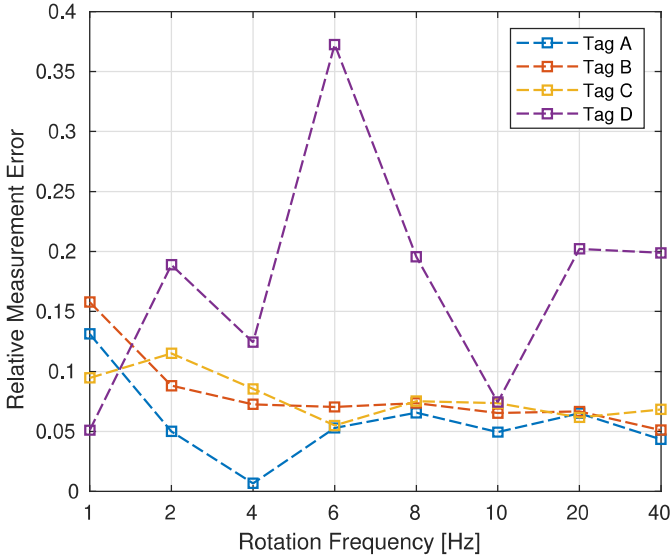

Figure 8: Impact of tag diversity on the mean relative measurement error.

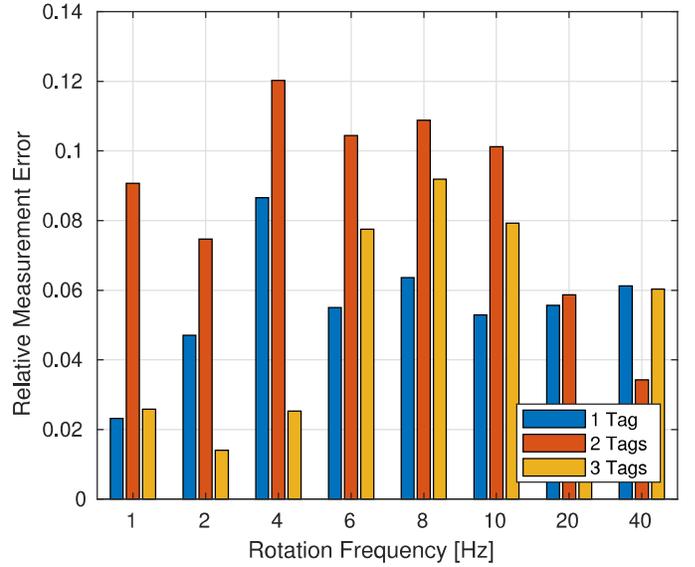

Figure 10: Relative measurement error as a function of the rotation frequency and the number of tags present.

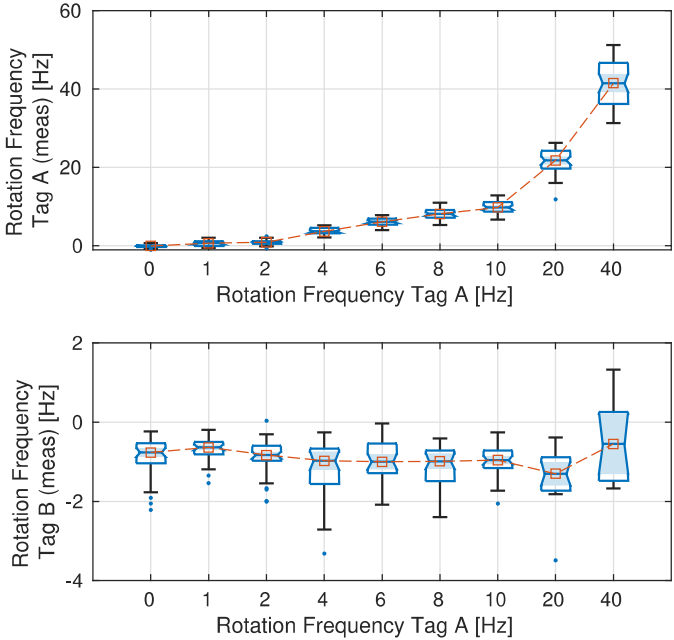

Figure 9: Absolute measurement error as function of the rotation frequency for 2 rotating tags.

E represents a double dipole antenna, resulting in a different polarization pattern than the other tags, which is expected to be the cause for the larger measurement error. Tag A, Tag B and Tag C show a decrease of the measurement error, with an increase in the rotation frequency, which matches the conclusions drawn in the previous section.

*4) Multiple Tags:* Besides measuring the rotation speed of a single tag rotating in the field of a reader, the proposed method allows for the acquisition of rotation data of more than one tag at a time. Every tag identified considering the EPC-C1G2 communication protocol has a unique ID. The acquired rotation frequency calculated from the corresponding backscatter signal is always assigned to this ID, thus, making it possible to obtain rotation data from more than one transponder within the range of the interrogator's antenna. Figure 9 shows the data obtained from positioning 2 tags in front of the antenna. The speed of Tag A was increased from static to $40\,\mathrm{Hz}$, while Tag B was rotating at a constant speed of $-1\,\mathrm{Hz}$, where the sign denotes that the rotation direction was counter-clockwise. It is visible that the different rotation directions were identified correctly and the assignment of the measurements to either of the ID's was correct for every measurement. 50 identifications were evaluated for each of the transponders in each measurement, except for the rotation speeds $20\,\mathrm{Hz}$ and $40\,\mathrm{Hz}$, where Tag B was identified less often. This decrease of the number of identifications results in a larger uncertainty of the measurement and is expected to have occurred due to the SDR reader platform used, where the maximum achievable output power is $20\,\mathrm{dBm}$.

An additional set of measurements was performed, where an additional, static Tag C was added to the previous scenario. Figure 10 shows a barplot of the mean relative measurement errors of Tag A at different rotation speeds for the 3 scenarios, where 1, 2 or 3 transponders were present in the field of the interrogator. It can be seen that the mean measurement error shows no significant discrepancies depending on the number of tags present. A larger measurement error can again be detected at lower rotation speeds which further proves the hypothesis, that at slower rotation, the angle covered during the measurement time is not sufficient for a more accurate determination of the rotation speed.

## VI. LIMITATIONS AND OUTLOOK

Given the proposed platform, consisting of an SDR UHF RFID reader and COTS UHF RFID tags, the main limitation is the maximum output power of the USRP N210 of $20\,\mathrm{dBm}$.

                                                                                    8

The signal phase considering polarization has a 180 degree ambiguity (rotation angle) if considering a perfect dipole tag antenna, which handicaps the use for tilt angle sensing. However, using tag antennas with a not solely linear polarization could allow for a full 360 degree rotation angle evaluation.

The measurement results presented in the previous section show that for a rotation speed between 0 and 40 Hz, the measurements are feasible. Given the used motor, larger rotation speeds were not achievable. Nevertheless, the proposed methodology should have increased accuracy where the rotation period matches the time of the backscatter message, which considering the above mentioned parameters would be at a frequency of approx 300 Hz. One limiting factor of the system at higher frequencies, using an SDR reader is expected to be the correct decoding of the tags backscatter response.

## VII. CONCLUSION

This work presents a methodology for real-time monitoring of the rotation frequency of a COTS UHF RFID tag based on the tag-reader polarization diversity. This is achieved using a platform consisting of an SDR UHF RFID reader and COTS UHF RFID tags. The main advantages of the presented approach are the fast acquisition rate of the rotation data, as well as the platform's ability to simultaneously measure the rotation speed of multiple UHF RFID tags rotating in the field of the reader. Experimental results ptove both the accuracy and the flexibility of the proposed methodology. Considering a sample rate of approximately $0.5\,\mathrm{ms}$, the mean measurement error at a rotation frequency of $10\,\mathrm{Hz}$ is $0.1\,\mathrm{Hz}$. The number of rotating tags present in the reader's field does not alter the measurement accuracy of the proposed platform, which represents a major advantage compared to the traditional rotation sensing approaches.

## REFERENCES

[1] Y. Song, F. R. Yu, L. Zhou, X. Yang, and Z. He, "Applications of the Internet of things (IoT) in smart logistics: A comprehensive survey," *IEEE Internet of Things Journal*, 2020.

[2] X. Jia, Q. Feng, T. Fan, and Q. Lei, "RFID technology and its applications in Internet of Things (IoT)," in *2012 2nd international conference on consumer electronics, communications and networks (CECNet)*. IEEE, 2012, pp. 1282–1285.

[3] K. Finkenzeller, *RFID Handbook: Fundamentals and Applications in Contactless Smart Cards, Radio Frequency Identification and Near-Field Communication*. John Wiley & Sons, 2010.

[4] "EPC™ Radio-Frequency Identity Protocols Generation-2 UHF RFID Standard."

[5] H. Landaluce, L. Arjona, A. Perallos, F. Falcone, I. Angulo, and F. Muralter, "A review of IoT sensing applications and challenges using RFID and wireless sensor networks," *Sensors*, vol. 20, no. 9, p. 2495, 2020.

[6] E. of Mathematics, "Rotation," https://encyclopediaofmath.org/index.php?title=Rotation.

[7] L. Yang, Y. Li, Q. Lin, H. Jia, X.-Y. Li, and Y. Liu, "Tagbeat: Sensing Mechanical Vibration Period With COTS RFID Systems," *IEEE/ACM transactions on networking*, vol. 25, no. 6, pp. 3823–3835, 2017.

[8] C. Duan, L. Yang, Q. Lin, Y. Liu, and L. Xie, "Robust Spinning Sensing with Dual-RFID-Tags in Noisy Settings," *IEEE Transactions on Mobile Computing*, vol. 18, no. 11, pp. 2647–2659, 2018.

[9] P. Castellini, M. Martarelli, and E. P. Tomasini, "Laser Doppler Vibrometry: Development of advanced solutions answering to technology's needs," *Mechanical systems and signal processing*, vol. 20, no. 6, pp. 1265–1285, 2006.

[10] K. Otsuka, K. Abe, J.-Y. Ko, and T.-S. Lim, "Real-time nanometer-vibration measurement with a self-mixing microchip solid-state laser," *Optics letters*, vol. 27, no. 15, pp. 1339–1341, 2002.

[11] P. Cheng, M. S. M. Mustafa, and B. Oelmann, "Contactless rotor RPM measurement using laser mouse sensors," *IEEE Transactions on Instrumentation and Measurement*, vol. 61, no. 3, pp. 740–748, 2011.

[12] M. Pedley, "Tilt Sensing Using a Three-Axis Accelerometer - Freescale Semiconductor Application Note," https://www.nxp.com/files-static/sensors/doc/app_note/AN3461.pdf.

[13] S. M. Seitz and C. R. Dyer, "View-invariant analysis of cyclic motion," *International Journal of Computer Vision*, vol. 25, no. 3, pp. 231–251, 1997.

[14] I. Laptev, S. J. Belongie, P. Pérez, and J. Wills, "Periodic motion detection and segmentation via approximate sequence alignment," in *Tenth IEEE International Conference on Computer Vision (ICCV'05) Volume 1*, vol. 1. IEEE, 2005, pp. 816–823.

[15] P. V. Nikitin, R. Martinez, S. Ramamurthy, H. Leland, G. Spiess, and K. Rao, "Phase Based Spatial Identification of UHF RFID Tags," in *2010 IEEE International Conference on RFID (IEEE RFID 2010)*. IEEE, 2010, pp. 102–109.

[16] S. J. Thomas, E. Wheeler, J. Teizer, and M. S. Reynolds, "Quadrature Amplitude Modulated Backscatter in Passive and Semipassive UHF RFID Systems," *IEEE Transactions on Microwave Theory and Techniques*, vol. 60, no. 4, pp. 1175–1182, 2012.

[17] R. Miesen, A. Parr, J. Schleu, and M. Vossiek, "360 Carrier Phase Measurement for UHF RFID Local Positioning," in *2013 IEEE International Conference on RFID-Technologies and Applications (RFID-TA)*. IEEE, 2013, pp. 1–6.

[18] M. Scherhaeufl, M. Pichler, E. Schimbaeck, D. J. Mueller, A. Ziroff, and A. Stelzer, "Indoor localization of passive UHF RFID tags based on phase-of-arrival evaluation," *IEEE Transactions on Microwave Theory and Techniques*, vol. 61, no. 12, pp. 4724–4729, 2013.

[19] C. Li, E. Tanghe, D. Plets, P. Suanet, N. Podevijn, J. Hoebeke, E. De Poorter, L. Martens, and W. Joseph, "Phase-based Variant Maximum Likelihood Positioning for Passive UHF-RFID Tags," in *2020 14th European Conference on Antennas and Propagation (EuCAP)*. IEEE, 2020, pp. 1–5.

[20] C. Li, E. Tanghe, D. Plets, P. Suanet, J. Hoebeke, E. De Poorter, and W. Joseph, "ReLoc: Hybrid RSSI-and Phase-Based Relative UHF-RFID Tag Localization With COTS Devices," *IEEE Transactions on Instrumentation and Measurement*, vol. 69, no. 10, pp. 8613–8627, 2020.

[21] M. C. Caccami, S. Manzari, and G. Marrocco, "Phase-Oriented Sensing by Means of Loaded UHF RFID Tags," *IEEE Transactions on Antennas and Propagation*, vol. 63, no. 10, pp. 4512–4520, 2015.

[22] X. Wang, J. Zhang, Z. Yu, S. Mao, S. C. G. Periaswamy, and J. Patton, "On Remote Temperature Sensing Using Commercial UHF RFID Tags," *IEEE Internet of Things Journal*, vol. 6, no. 6, pp. 10715–10727, 2019.

[23] X. Lai, Z. Cai, Z. Xie, and H. Zhu, "A novel displacement and tilt detection method using passive UHF RFID technology," *Sensors*, vol. 18, no. 5, p. 1644, 2018.

[24] Voyantic, "Voyantic Field Engineer Kit," http://voyantic.com/products/tagformance-pro/accessories/field-engineer-kit, accessed on 2022-01-01.

[25] C. Jiang, Y. He, X. Zheng, and Y. Liu, "Orientation-aware RFID tracking with centimeter-level accuracy," in *2018 17th ACM/IEEE International Conference on Information Processing in Sensor Networks (IPSN)*. IEEE, 2018, pp. 290–301.

[26] S. Pradhan, S. Li, and L. Qiu, "Rotation Sensing Using Passive RFID Tags," in *Proceedings of the Twenty-second International Symposium on Theory, Algorithmic Foundations, and Protocol Design for Mobile Networks and Mobile Computing*, 2021, pp. 71–80.

[27] L. Shangguan and K. Jamieson, "Leveraging electromagnetic polarization in a two-antenna whiteboard in the air," in *Proceedings of the 12th International on Conference on emerging Networking EXperiments and Technologies*, 2016, pp. 443–456.

[28] I. Toyoda, "Polarization Modulation," in *Modulation in Electronics and Telecommunications*. IntechOpen, 2019.

[29] A. Gerrard and J. M. Burch, *Introduction to matrix methods in optics*. Courier Corporation, 1994.

[30] N. Kargas, F. Mavromatis, and A. Bletsas, "Fully-coherent reader with commodity SDR for Gen2 FM0 and computational RFID," *IEEE Wireless Communications Letters*, vol. 4, no. 6, pp. 617–620, 2015.

[31] I. Impinj, "Application Note - Low Level User Data Support," https://support.impinj.com/hc/en-us/articles/202755318-Application-Note-Low-Level-User-Data-Support, 2013.

[32] K. S. Rao, P. V. Nikitin, and S. F. Lam, "Antenna design for uhf rfid tags: A review and a practical application," *IEEE Transactions on antennas and propagation*, vol. 53, no. 12, pp. 3870–3876, 2005.